\documentstyle[12pt,aasms4]{article}

\def\reference{\par\noindent\hangindent=1cm\hangafter=1}

\newcommand{\eq}{\begin{equation}}
\newcommand{\ee}{\end{equation}}

\def\t0{\theta_{\circ}}

\def\be{\begin{equation}}
\def\en{\end{equation}}

\def\gapp{\ \lower 3pt\hbox{${\buildrel > \over \sim}$}\ }
\def\lapp{\ \lower 3pt\hbox{${\buildrel < \over \sim}$}\ }

\lefthead{Jiang \& Ip}
\righthead{Planetary System}

\begin{document}

\title{The Planetary System of Upsilon Andromedae}

\author{Ing-Guey Jiang \& Wing-Huen Ip}

\affil{Academia Sinica, Institute of Astronomy and Astrophysics, Taipei, 
Taiwan\\
National Central University, Chung-Li, Taiwan}

\authoremail{jiang@asiaa.sinica.edu.tw}

\begin{abstract}
The bright F8 V solar-type 
star upsilon Andromedae has recently been reported to have 
a system of three planets of Jovian masses. 
In order to investigate the orbital stability and mutual gravitational
interactions among these extrasolar planets, both forward and backward 
integrations from the latest observed orbital elements for
all three planets' orbits  have been performed under the coplanar 
assumption. 
We reconfirm that the middle and the outer planet have strong 
interaction leading to large time variations in the eccentricities 
of these planets, which was shown by the previous studies. However, we
discuss the validity of the ignorance of the innermost planet.
We argue that this planetary system
is likely to be stable and oscillate around current orbital elements
since it was formed.

We suggest that one possible way to produce these
orbital elements: the innermost planet has very low eccentricity
but the outermost planet has high eccentricity could be the interaction 
with the protostellar disc.
 
\end{abstract}

\keywords{celestial mechanics - stellar dynamics - planetary system
}

\section{Introduction}

As a result of recent observational efforts, 
the number of known extrasolar planets increased dramatically. Among these 
newly discovered planetary systems, upsilon Andromedae
system appears to be the most interesting one 
because of the presence of three 
planetary members (Butler et al. 1999). 

Since the discovery of the planetary system of upsilon Andromedae, 
the dynamics of this multiple planetary system
with intriguing orbital configuration has drawn a lot of attention. 
Table 1 is the latest (as of 21st August, 2000) orbital elements 
obtained by 
the Marcy's group  
(http://exoplanets.org/esp/upsandb/upsandb.html).
It would be interesting  to understand
the origin of the orbital configurations 
of these three extrasolar planets and their mutual interactions.

\begin{table}[hbt]
  \caption[]
{\label{tab:sum}The Orbital Elements}
\renewcommand{\arraystretch}{1.2}
\begin{tabular}{ccccc}
\hline
Planet      & $M {\rm sin} i/M_{J}$ & a/AU & e & $\omega$ (deg)  \\
\hline
 B      & 0.69    & 0.059 & 0.01 & 316.4  \\

 C      & 2.06    & 0.827 & 0.23 & 247.2 \\

 D      & 4.10    & 2.56 & 0.35 & 250.6 \\
\hline
\end{tabular}
\end{table}

It is a remarkable fact that the eccentricities of the companion 
planets increase from 0.01 for the innermost member to a value 
as large as 0.35 for the outermost member. 
An interesting question is therefore if  orbital evolution
leads to this configuration.
This question is related to whether these planets 
interacted strongly in the past.

Without studying the past history, 
several groups have investigated this system by forward orbital 
integration from the observed orbital elements.
Laughlin \& Adams (1999) simplified
the model computation by ignoring the innermost planet. 
They found that the upsilon Andromedae system should experience 
chaotic evolution for all parameters derived from 
present observations.
In spite of the large amplitudes of the eccentricities of the middle 
and outer planets, this system could remain non-crossing over a time 
interval of  
2-3 Gyr for a significant number of the cases studies.

Rivera \& Lissauer (2000) did many extensive calculations for both 
nearly coplanar systems and mutually inclined orbits. For coplanar 
systems, they found that the nominal Lick data systems are more stable 
than the system with the nominal 
Advanced Fiber Optic Echelle (AFOE) parameters. 
They also explore different values of the 
overall mass factor $m_f = ({\rm sin} i)^{-1}$ and found that the systems with 
smaller $m_f$ are
more stable for both Lick data systems and AFOE data systems.

Rivera \& Lissauer (2000) also ignored the innermost planet for some 
calculations and they found that two-planet systems with AFOE 
parameters typically last much longer than their three-planet analogs.

In this paper, we focus on the most stable configuration in 
Rivera \& Lissauer (2000), i.e. the coplanar Lick data system with 
$m_f=1$. First of all, we study the orbital interaction between three planets 
by an forward integration of $10^5$ yrs. We analyze the interaction by 
comparing  the numerical result to the analytical equations.
The main goal of this analysis is to understand the validity of ignoring
the innermost planet.

In order to study the past history and 
origin of the orbital elements, we perform the
backward integration, which was not done in the previous work.
In order to test the correctness of 
our calculations, we also do forward integration and check if the results
are consistent with the results of Rivera \& Lissauer (2000).
We therefore do both forward and backward integrations of three-planet 
system for $10^6$ yrs. We also do 
both forward and backward integrations of two-planet 
system for $10^8$ yrs. 

From these calculations, we found that it is 
a good approximation to ignore the innermost planet 
of the coplanar Lick data system with 
$m_f=1$ for a long term integration. This was already shown in 
Rivera \& Lissauer's results of coplanar Lick data systems 
(Their three-planet system survived at least $10^8$ yrs and 
two-planet system survived at least $10^9$ yrs.)  
 What is new here is that we show that 
the results of backward integration behave similarly to the forward 
integration and therefore the system is likely to oscillate around
current orbital elements since its formation.
The origin of these orbital elements is complicated but we use one 
simple calculation of disc-planet interaction  to argue that 
the protostellar disc might be important in causing 
these orbital elements.

In Section 2, we describe the simulation model. We analyze the planet-planet 
interaction for a time scale of $10^5$ in Section 3. In Section 4, 
we investigate the dynamical origin of the orbital elements. 
We provide the conclusions in Section 5.

\section{The Simulation Models}

\subsection{The Orbital Integration}

We use mixed variable symplectic (MVS) integrator in the 
SWIFT package (Levison \& Duncan 1994) to integrate 
the orbit for the planetary system 
of upsilon Andromedae. The initial condition was from Table 1, which
is the latest orbital elements determined by Marcy's group.
We modified MVS integrator to be able to do backward integration. 
This can be done because of the integrator's symplectic property. 
We follow Rivera \& Lissauer (2000) to use 0.23 days as our timestep
for both forward and backward integration when all three planets
are included. We also follow Rivera \& Lissauer (2000) to use 
2.42 days as our timestep
for both forward and backward integration of two-planet system.

\subsection{The Planet-Disc Interaction}

In the calculation of planet-disc interaction, 
we assume both the mass of the central star 
and the gravitational constant $G$
to be unity. The masses of the 
planets are set to be zero, so there is no interaction between different
planets.

We include a disc into the Hermit integrator developed by
 Sverre Aarseth (Markino \& Aarseth 1992, Aarseth, Lin \& Palmer 1993) 
The disc has density profile as:
\begin{equation}
\rho (r,t) =   \left\{ \begin{array}{ll}
0& \mbox{$r \le \epsilon $} \\
c\ \ r^{-11/4} \exp(-t/\tau) & \mbox{$r> \epsilon$}.\end{array}
\right.
\end{equation}
We choose $\epsilon=0.01$, $c=0.001$ and $\tau = 30$ so that
the mass of the disc would be about 0.1 and the disc would be depleted
in a time scale of 30.

\section{The Planet-Planet Interactions}

The numerical integration of the three-planet system is expensive 
because the period of the innermost planet is only 4.617 days. 
Therefore, the innermost planet is usually ignored when one wishes to 
study long-term stability.

In order to see if the omission of the innermost planet 
is a good approximation,
we analyze the interaction between three planets by comparing
the numerical result of $10^5$ yrs to the analytic 
equations in this section.
 
The initial condition of the numerical integration was from Table 1. 
The time variations of the semi-major axises and eccentricities  
are given in Figure 1. Rather significant variations 
in eccentricities
are found which can be understood in terms of planet-planet interaction. 
As usually encountered in celestial mechanics, the semi-major 
axes remain nearly invariant while the eccentricities could follow 
rapid variations of large amplitudes as a consequence of angular momentum 
exchange.


We can understand the details of this
interaction from the results in Figure 1 with the help of equation 
(2.144) and equation (2.147) in Murry \& Dermott (1999):

\begin{equation}
\dot{C_i} = \frac{\mu_i}{2 a_i^2} \dot{a_i},
\end{equation}
where $C_i$ is the energy, $\mu_i=G(M_{\star}+M_i)$, $a_i$ is the semi-major 
axis. The index $i$ can be ${\rm m}$ for the middle planet 
and ${\rm o}$ for the outer planet. Therefore, $M_{\rm m}$ is the mass
of the middle planet and $M_{\star}$ is the mass of the central star.
  
\begin{equation}
\frac{de_i}{dt} = \frac{e_i^2-1}{2e_i}[2\dot{h_i}/h_i + \dot{C_i}/C_i],
\end{equation}
where $e_i$ is the eccentricity and $h_i$ is the angular momentum.
  
The main result from the numerical 
simulation is that the semi-major 
axis of the middle planet 
does not change much but the eccentricity change quickly 
between 0.125 and 0.225 .  This tells us that $\dot{C_{\rm m}}$ 
is small from Equation (2).
Thus, from Equation (3), we know that all the quick eccentricity 
variation is due to the angular momentum variation. Bottom panel of 
Figure 1 showed that 
the frequencies of the eccentricity variations of the middle and outer planet 
are very close and thus the angular momentum change of the middle 
planet must be due to the forcing from the outer planet. The 
semi-major axis of the outer planet also has certain variation and this 
should be from the energy exchange with the middle planet because 
Equation (2) tells us that 
the time derivative of semi-major axis is related to the 
time derivative of energy.
The reason why the variation of semi-major axis of the middle planet
looks so small is due to the form of Equation (2): For the same $\dot{C_i}$,
$\dot{a_i}$ is smaller for smaller $a_i$.
The reason why the variation of $e_i$ for the outer planet is smaller is 
partially because $(e_i^2-1)/(2e_i)$ for the outer is about 1.05
but 2.4 for the middle planet in Equation (3).
Therefore, the outer and middle planets are indeed interacting strongly.

On the other hand, from the bottom panel of Figure 1, 
we see that the frequency of 
eccentricity variation of the innermost planet is very different from 
the other two planets and the semi-major axis is almost constant. The 
innermost planet does not involve that much of the dynamics of the 
middle and outer planets. Therefore, it is a good approximation to ignore 
the innermost planet when one needs to do it for a long-term integration.

  
%
%

\section{The Origin of Orbital Elements}

The planetary system of upsilon Andromedae is interesting not only
because of the presence of multiple planets but also because of 
the orbital configuration.

It is therefore important to investigate if orbital evolution
leads to this current configuration:
the innermost planet has very low eccentricity but the outermost
planet has much higher eccentricity. 

There are two obvious ways to lead to the current orbital elements
of the exoplanets of upsilon Andromedae.
One way is that even all of these three planets had similarly 
small  eccentricities when they were form, the long-term orbital evolution can
cause the outer two planets to have higher eccentricities.
Another way is that, these orbital configuration was originally due to the
disc-planet interaction when the system was formed. 
After that, the orbital elements are kept to be similar but 
oscillate around the current values.

We can investigate the first possibility  by doing backward integrations.
In order to check if our numerical code can produce the results which
are consistent with the results of Rivera \& Lissauer (2000), we also do 
the usual forward integrations.

Figure 2 are the results of both backward and forward integrations for
$10^6$ yrs. The semi-major axises of three planets are almost constant
all the time except the small fluctuations for the outermost planet.
However, the eccentricities of all three planets oscillate in very large 
amplitudes: The innermost planet oscillate between 0 and 0.15, the middle
planet oscillate between 0.125 and 0.225, the outermost planet oscillate
between 0.33 and 0.35 . Therefore the orbital elements remain unchanged 
but just oscillate around the current observed values for both backward 
and forward integrations.

To check if this is still the case for a longer time scale, we integrate
100 times longer, i.e. $10^8$ yrs. 
We ignore the innermost planet in this case
because the long-term integrations are far more expensive and from the last 
section we know that the ignorance of the innermost planet is a good 
approximation.

Figure 3 are the results of both backward and forward integrations for
$10^8$ yrs. The results are just the extension of the results in Figure 2, so
the orbital elements still oscillate around the current observed values 
for both backward and forward integrations of $10^8$ yrs.  
From Figure 3, we can see that it is already impossible to see the lines
in the figure. We also integrate for a longer time scale (order of Gyr). 
The result
remains to be the same and it is not necessary to 
provide this figure in the paper because it looks almost the same as 
Figure 3.

From the above calculations of forward integrations, it is encouraging 
that we can produce 
the results which are consistent with those of 
Rivera \& Lissauer (2000).  
From the calculations of backward integrations, 
we know that the long-term orbital evolution
might not be able to produce the current orbital configuration because the 
orbital elements do not really change but just oscillate around the current
values.
Thus, we should test if the orbital configuration was originally due to the
planet-disc interaction when the system was formed. 
After the formation process, the orbital elements keep to be similar but 
oscillate around the current values.

We use the model we described in Section 2.2 to simulate the planet-disc 
interaction. We assume the inner planet is at $r=0.8$ and the outer planet
is at $r=2.5$ and both of them have eccentricity $e=0.2$ initially.
Figure 4 is the result for time evolution of both semi-major axis and 
the eccentricity.

Because of the interaction with the disc, both eccentricities increase
and oscillate. The existence of the disc makes the gravitational force
experienced by the planets more complicated and increase the eccentricities.
(If there were energy dissipation during the planet-disc interaction,
the eccentricities of both planets should decrease.) 
 However, the outer planet's eccentricity is pumped to
a higher value. During the depletion of the disc, the  eccentricities
of both planets gradually settle down to a stable value and stop oscillation.
The final eccentricity of the outer planet is higher than the final 
eccentricity of the inner planet. 
Further, 
the final semi-major axes of both planets are about the same as initial
values. Therefore, 
we have produced a orbital configuration that the outer planet
has higher eccentricity than the inner planet by the interaction with the 
disc. 

We found that the results are qualitatively about the same 
when we explore different values of parameters. Though we make 
the value of $\tau$
to be very small to save our computational time, 
it would not affect the final values of
eccentricities.

\section{Conclusions}

The dynamics of extrasolar planetary systems continues to be a fascinating 
and important subject. One of the very intriguing aspects for 
the extrasolar planets is the existence of orbits of high eccentricities. 
The dynamical cause of high eccentricities is unknown but should be 
related to the early history of planetary formation or stellar encounters.

Among these discovered extrasolar planetary systems, 
the upsilon Andromedae planetary system appears to be one of the most 
interesting one because of the presence of multiple planets and also
the orbital configuration.
Even though the cause of the high eccentricity of the outer planet 
of the upsilon Andromedae planetary system
is not clear yet, our computation showed that the gravitational 
interaction of the
four-body system is potentially very complex and the 
eccentricity of the middle planet oscillate between 0.125 and 0.225 over 
the time interval of orbital integration. The middle and the outer planets
are indeed interacting strongly and the ignorance of the innermost
planet can be a good approximation for the long-term integrations.

On the other hand, one general observational fact is that extrasolar 
planets with high eccentricities usually have larger semi-major axes 
than those with small eccentricities.  
The upsilon Andromedae planetary system follows this observational trend. 

We investigate the origin of the current observed orbital elements 
of the upsilon Andromedae planetary system. Our long-term 
backward integration (order of Gyr) shows that the current orbital 
configuration was not caused by the orbital evolution because 
the orbital elements do not really change during the backward integration
but just oscillate around the current values.  

Our results show that the interaction between the exoplanets and 
the protostellar disc might
lead to the current orbital configuration.
It is possible that the model will be even more pertinent if there
is energy dissipation when the planets interact with 
 the disc. This energy dissipation might make the eccentricity of 
the inner planet decrease more than the outer planet would 
because there could be more dissipation around the inner disc.
In this case, the eccentricity difference between two planets
might be even larger. We hope to come back to this issue in the future.
    
\section*{Acknowledgements}
Ing-Guey Jiang wishes to take this opportunity to 
acknowledge James Binney's supervision on
Dynamics during his study at Oxford and also the hospitality
of Doug Lin and Martin Duncan during his visits at both
UC Santa Cruz and Queen's University.

\section*{REFERENCES}
\begin{reference}

\reference Aarseth, S.J., Lin, D.N.C. \& Palmer, P.L., 1993, ApJ, 403, 351 

\reference Butler et al., 1999, ApJ, 526, 916.  
\reference Laughlin, G. \& Adams, F.C., 1999, ApJ, 526, 881.

\reference Levison, H. F. \& Duncan, M. J., 1994, Icarus, 108, 18

\reference Makino, J. \& Aarseth, S.J., 1992, PASJ, 44, 141

\reference Murray, C. D. \& Dermott, S. F., 1999, Solar System Dynamics, 
Cambridge University Press, Cambridge. 

\reference Rivera, E. J. \& Lissauer, J. J., 2000, ApJ, 530, 454. 

\end{reference}

\clearpage

\begin{figure}[tbhp]
\epsfysize 7.0in \epsffile{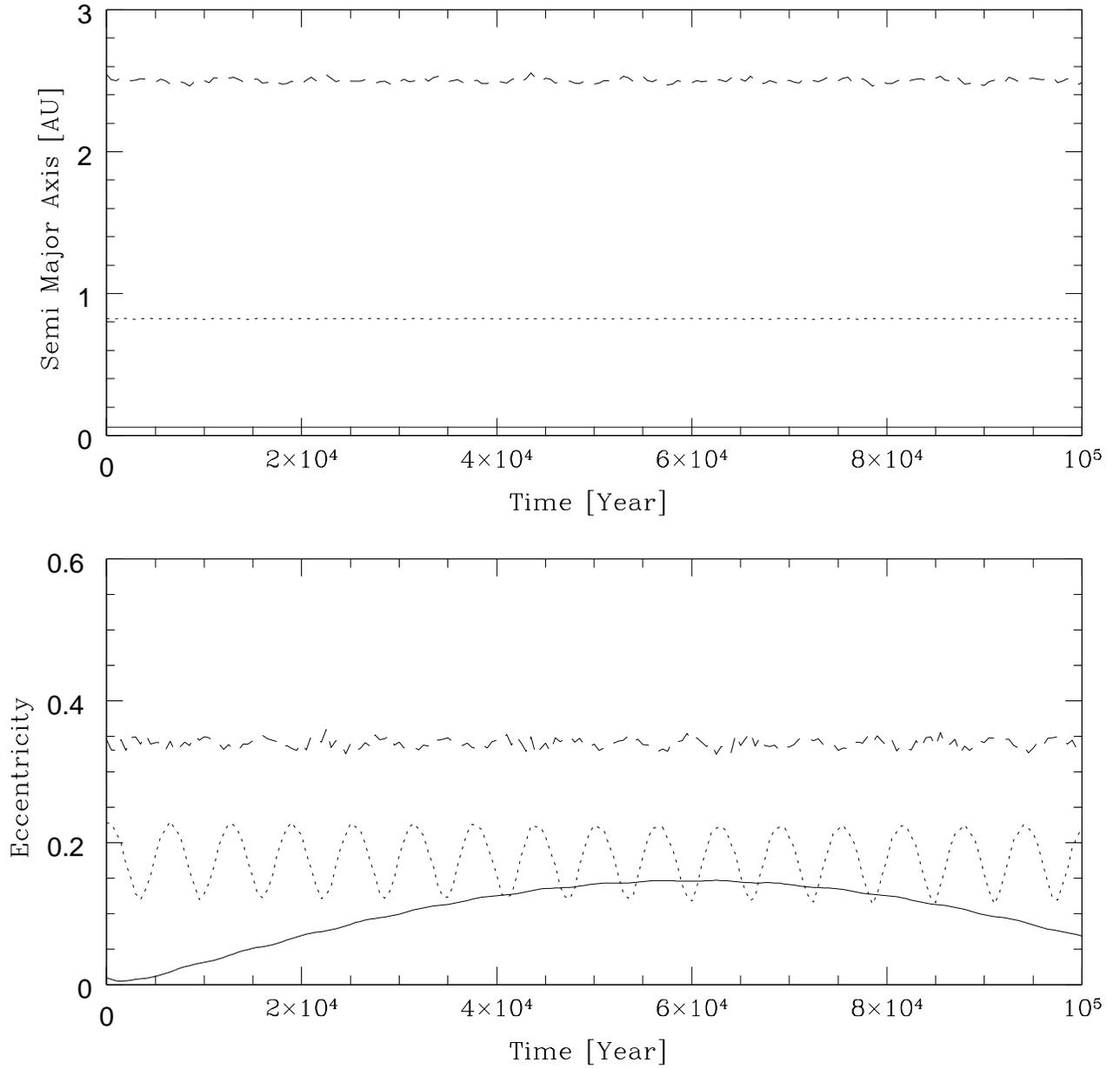}
\caption{
The semi-major axes (top panel) and the eccentricities 
(bottom panel) of all planets as function of time, where 
the solid line is for the inner planet, the dotted line is for 
the middle planet and 
the dashed line is for the outer planet.
}
\end{figure}

\begin{figure}[tbhp]
\epsfysize 7.0in \epsffile{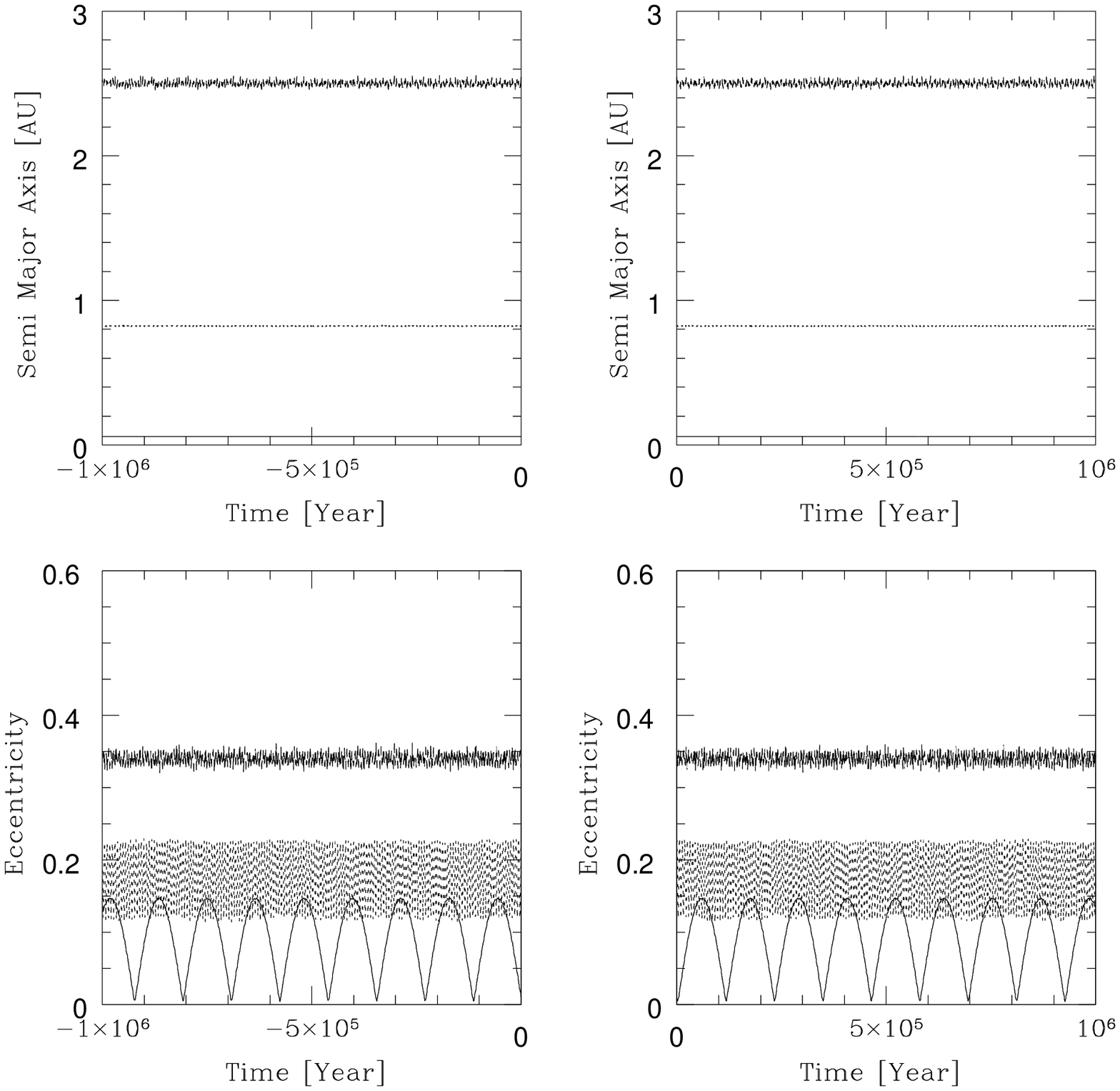}
\caption{
The semi-major axes (top panels) and the eccentricities 
(bottom panels)
of all planets as function of time for both
backward (left panels) and forward (right panels) integrations, 
where the solid line is for 
the inner planet, the dotted line is for the middle planet and 
the dashed line is for the outer planet.
}
\end{figure}

\begin{figure}[tbhp]
\epsfysize 7.0in \epsffile{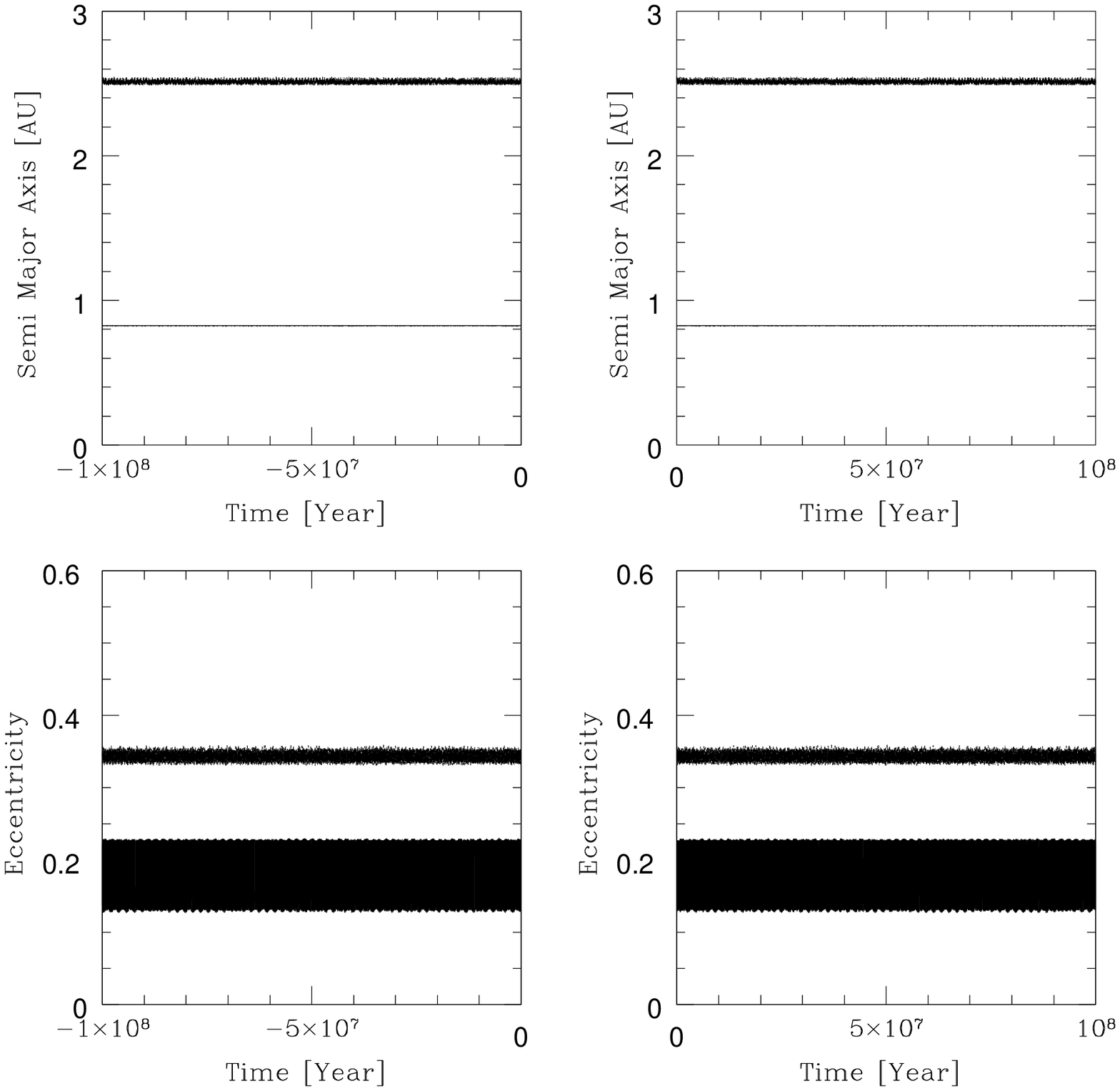}
\caption{
The semi-major axes (top panels) and the eccentricities 
(bottom panels)
of planets as function of time for both
backward (left panels) and forward (right panels) integrations, 
where the solid line is for 
the middle planet, the dotted line is for the outer planet.
}
\end{figure}

\begin{figure}[tbhp]
\epsfysize 7.0in \epsffile{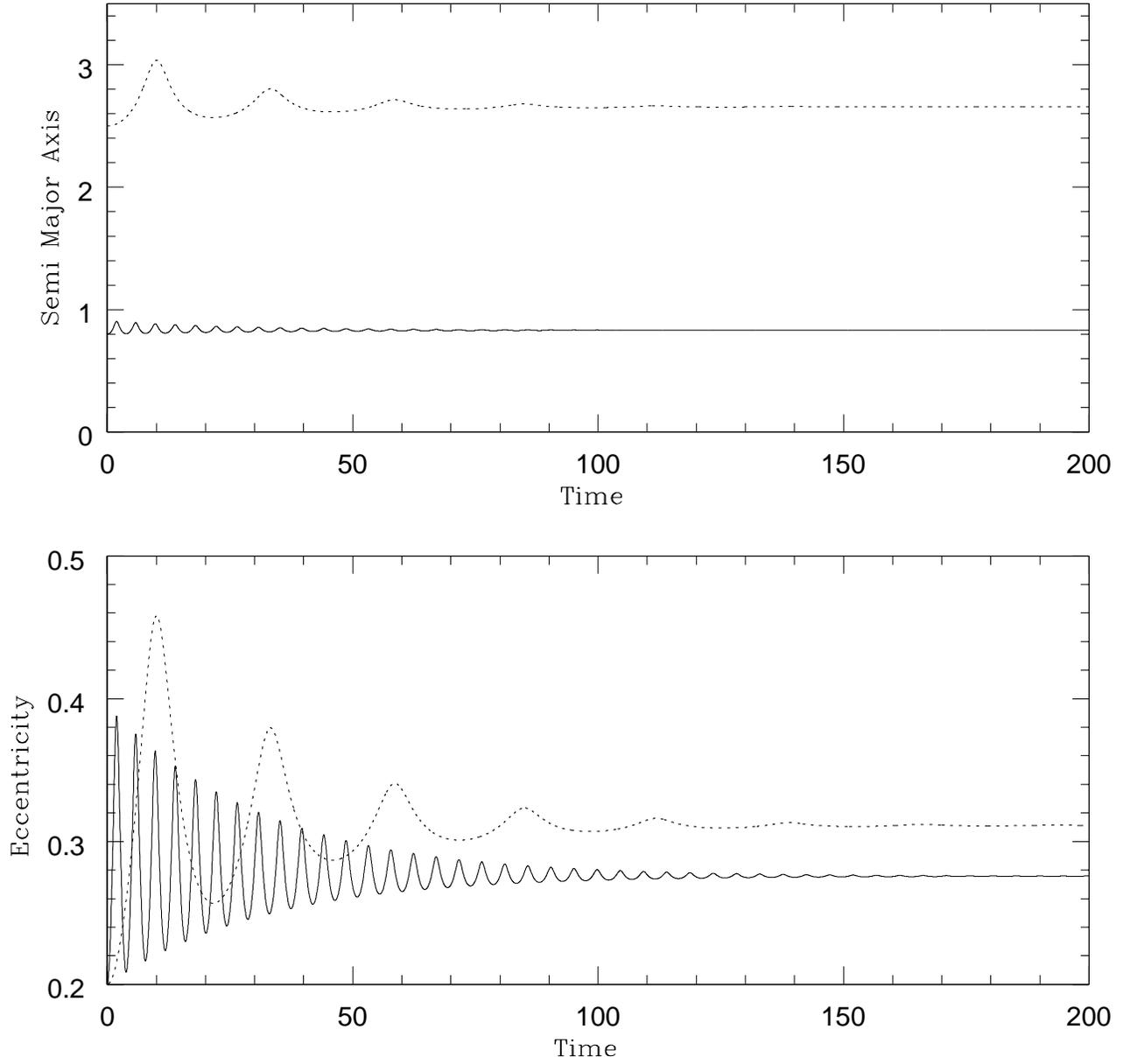}
\caption{
The semi-major axes (top panel) and the eccentricities 
(bottom panel) of planets (under the interaction 
with the disc) as function of time, where 
the solid line is for the inner planet, the dotted line 
is for the outer planet.
}
\end{figure}

\end{document}